\documentclass[aps,pre,floats,preprint]{revtex4}

\usepackage{amsfonts}
\usepackage{times}

\usepackage{graphicx}
\renewcommand{\a}{\alpha}

\newcommand{\e}{\epsilon}

\renewcommand{\r}{\rho}
\renewcommand{\L}{\Lambda}

\renewcommand{\r}{\rho}
\renewcommand{\b}{\beta}
\renewcommand{\l}{\lambda}
\renewcommand{\inf}{\infty}

\begin{document}
\title{Numerical methods for fluctuation driven interactions between dielectrics}
\author{S. Pasquali$^1$, F. Nitti$^2$, A. C. Maggs$^1$}
\affiliation{$^1$Laboratoire de Physico-Chime Th\'eorique, UMR CNRS-ESPCI
  7083, 10 rue Vauquelin, 75231 Paris Cedex 05, France.  \\ $^2$CPHT, Ecole
  Polytechnique, 91128, Palaiseau, France ( UMR du CNRS 7644).}
\date \today 

\begin{abstract}
  We develop a discretized theory of thermal Casimir interactions to
  numerically calculate 
the interactions between fluctuating dielectrics.  From a
  constrained partition function we derive a surface free energy, while
  handling divergences that depend on system size and discretization.  We
  derive analytic results for parallel plate geometry in order to check the
  convergence of the numerical methods.  We use the method to calculate
  vertical and lateral Casimir forces for a set of grooves.
\end{abstract}
%\pacs{1.234}
\maketitle
\section{Introduction}

Dispersion forces are long range interactions due to thermal or quantum
fluctuation fields. The first theoretical derivation was due to Casimir
\cite{Casimir} who predicted the attraction of two neutral plates. A general
continuum theory, developed by Lifschitz, \cite{Lifshitz} addresses the
non-additivity of these forces and is often used to interpret experiments
\cite{Tabor}. Calculation of dispersion forces requires knowledge of the
frequency dependent dielectric constant; information which is accessible from
spectroscopy.  The results of the theory can be expressed as a sum over
Matsubara frequencies, \cite{Milton}.  However, one can isolate the zero
Matsubara frequency and recognize that the corresponding contribution is both
temperature dependent and independent of $\hbar$; it is purely classical and
depends on the static dielectric constant, $\epsilon(\omega=0)$.  The forces
derived from all other frequencies depend on dynamic dipole fluctuations and
require information on the frequency dependence of the dielectric
permittivity.  
For biophysical systems, mainly composed of water and lipids, Ninham and
Parsegian \cite{Parsegian2,Parsegian} have shown that zero frequency
gives a contribution to the interaction which is at least as important as the
interactions coming from the UV region of $\epsilon(\omega)$. 
This feature makes biological materials rather unique, and justifies the study
of the zero frequency contribution alone.
Zero Matsubara frequency forces go under different names
including static van der Waals forces \cite{Netz}, classical thermal Casimir
forces \cite{Dean1}, and Keesom interactions.  For the simplest geometries one can
calculate the free energy analytically \cite{Kampen,Ninham}.

For more complex geometries this problem has been 
approached from various directions, both in its full quantum formulation, 
and in the high temperature, classical, regime.
The range of techniques used is very wide, including Green's function formulations 
\cite{Emig1}, world-line numerics \cite{Gies}, and path-integral formulations 
\cite{Feinberg}. These techniques have allowed for great progress in linking the theory 
with experimental observation, but, as of now, no general method is avaliable
for arbitrary geometries and arbitrary values of the parameters.
%Our numerical approach targets any discretizable geometry in
%high dielectric contrast, a sector which is not easily accessible by the other
%methods.
%none of them can give a general description 
%of the phenomenon, valid for arbitrary geometries and in all parameters range.

Recently, there has been a renewed interest in Casimir forces among soft condensed matter 
physicists
as it was recognized that these forces can play an important role in
biophysical systems \cite{Dean1, Netz}. %  For highly polarizable materials,
% including water, the static van der Waals force can be more important
% than the quantum van der Waals force when the optical properties
% of the bodies in contact are very similar.
Ninham and Parsegian observations have opened the road for new formulations of dispersion forces
that focus exclusively on their classical part. 
Dean and Horgan showed how to
calculate the free energy using a classical partition function without 
the machinery of Matsubara frequencies \cite{Dean1}.  This formulation has the
advantage of making the physics more transparent, as thermal effects are
considered explicitly from the start, rather than obtained as a limiting case
of a quantum theory.

In order to address a wider variety of geometries than is
possible analytically, we introduce numerical methods that can be used to study
thermal Casimir forces on a lattice. In particular we address the nature of
the various terms appearing in the free energy of a dielectric system and
study their variation with the discretization.  This corresponds to the issue
of regularization of divergences in a continuous infinite system.  We apply
our methods to a set of rectangular and sinusoidal grooves in order to characterize both
longitudinal and transverse Casimir interactions.

In this paper we are interested in computing the free energy due to thermal
fluctuations of the electrostatic field between dielectric bodies.  In this
classical perspective, as noted above, the dielectric permittivity $\e$ is
taken to be a function of space, $\e(r)$, but not a function of frequencies,
i.e. $\e(r,\omega=0)$.  Our main goal is to develop a formalism which is a
suitable starting point for efficient and versatile numerical methods, which
can be applied to arbitrary geometries.  Considering the classical,
time-independent regime alone, allows us to present the general formulation
and the numerical approaches on the simplest system.  After a calibration
  which is run with convenient but non-physical values of $\e(r)$, we focus on
  systems with physically relevent material properties, that is systems for
  which the optical properties of the object immersed in water match those of
  water. This is the case if we choose a material whose refractive index is
  similar to that of water, $n\sim 1.3$, giving a dielectric constant for the
  material of $\e_{mat}=1.7$, while $\e_{H_2O}=80$.

In separate works we extend our approach to quantum regimes, considering the
full dielectric function $\e(r,\omega)$ \cite{PRL}.  We note also the
  existence of a second regime where thermal interactions can dominate even in
  the absence of matching of optical properties \cite{Daicic}; separations $L$
  satisfy $L >> \hbar c/(2 k_B T)$. This case in only relevant when $L>> 4\mu
  m$ and will not be discussed further here.

The paper is organized as follows: In section~\ref{sec_lattice} we discretize
a set of dielectrics and derive the main theoretical results; in
section~\ref{numerics} we study the discretized system using matrix
diagonalization and Monte Carlo simulation.  In section~\ref{analytics} we
present a continuum calculation which derives the free energy of a single
homogeneous slab with periodic boundary conditions. This result is compared to
the numerical results in section~\ref{results}, where we also show how the
numerical methods can be used to calculate the free energy for grooved
surfaces.

\section{Lattice Formulation} 
\label{sec_lattice}

The free energy of a dielectric in the absence of free charges is
calculated from the partition function \cite{Tony-Ralph}
\begin{equation}
\label{Z} \mathcal{Z} = \int\mathcal{D}[{\bf D}]
\;\prod_r\delta(\nabla\cdot{\bf D})\, e^{-\b\mathcal{U}}, 
\end{equation}
with ${\bf D}({\bf r})$ the electric displacement.  
The Boltzmann factor is the electrostatic energy in the presence of a dielectric medium
\begin{math}
   \mathcal{U}=\int \left( {{\bf D}^2({\bf r})\, 
}/ {2\,\e({\bf r})}
\; \right) d^3{\bf r}
\end{math}.  Gauss' law $\nabla\cdot{\bf D}=0$ is imposed in (\ref{Z}) as a
constraint.  In the continuum limit, and also in infinite volume,
eq.~(\ref{Z}) is ill-defined, as it contains divergences. We thus discretize
in order to find an unambiguous definition of the free energy. We will then
remove contributions to the free energy that diverge when the system size
diverges, or mesh size is taken to zero, in order to calculate the remaining
long ranged contributions to the free energy.

We use a cubic lattice of $N$ nodes and $3N$ links.
We use the word lattice to refer to a regular grid.
Since we are discretizing a macroscopic theory our lattice is not related to a
physical atomic lattice. To be meaningful the lattice spacing should be large
compared to atomic dimensions (so that one can discuss electrostatic
interactions in terms of a dielectric constant) but small compared to the
physical system at hand.

For simplicity of notation we set the lattice spacing equal to $1$.
% This choice is irrelevant to the final results we will obtain since, due to
% the scaling of the free energy with the area and the distance, the units in
% which lengths are measured cancel out.
Physical results depend on the ratio between the size of the bodies and 
their separation. 
Dimensional information can be reintroduced by standard scaling arguments.
Physical quantities are associated with nodes or links; scalars ``live'' on nodes, vectors on links.
Thus $D_{nx}$ is assigned to the link leaving the node $n$ in the positive $x$
direction. The discretized divergence $\nabla\cdot {\bf D}$ gives the total
flux at a lattice site.  The dielectric permittivity $\e$ is also associated
with the links so that the discretized energy density takes the form
\begin{math}
\frac{D_x^2}{2\e_x}+\frac{D_y^2}{2\e_y}+\frac{D_z^2}{2\e_z}
\end{math}.  On a lattice, eq.~(\ref{Z}) becomes
\begin{equation} \label{Z_deltaD}
\mathcal{Z} = \int\mathcal{D}[{\bf D}]
\left(\;\prod_{nodes}^{N-1}\delta(\nabla\cdot{\bf D})\right)
\exp\left[-\frac{\beta}{2}\sum_{links} \frac{D_l^2}{\e_l} \right].  
\end{equation}
where $\e_l$ is the permittivity associated with the link $l$.  Notice that,
in periodic boundary conditions we impose $N-1$ delta-functions; the
constraint on the $N^{\mathrm{th}}$ node is automatically satisfied when $N-1$
are imposed since
\begin{math}
\int d^3{\bf r} \; {\nabla \cdot} {\bf D} =0
\end{math}

We impose Gauss' law by introducing an auxiliary scalar field $\phi
\equiv\{\phi_n\}_{n=1\ldots N}$ as a Lagrange multiplier.
It was shown in \cite{Tony} that $\phi$ is the static electrostatic potential.
Using the identity
\begin{equation} \label{lagrange_m}
(2\pi)^N\prod_{nodes} \delta({\bf \nabla}\cdot {\bf D}) = \int
\mathcal{D}[\phi]\, \exp\left[-i \sum_{nodes}\phi{\bf \nabla}\cdot {\bf
    D}\right].  
\end{equation}
This identity involves the product over all $N$ nodes, so we must introduce an
extra constraint $\delta(\phi_N)$ to remove the integral over $\phi_N$.
Dropping irrelevant prefactors. Eq.~(\ref{Z_deltaD}) becomes 
\begin{equation}
 \label{Z_phi} \mathcal{Z} =
\int\mathcal{D}[\phi]\mathcal{D}[{\bf D}]\; \exp\left[-\frac{\beta}{2}
  \sum_{links} \frac{D_l^2}{\e_l}\right]\;\delta(\phi_N)\,
\prod_{nodes}^{N}\exp\left[-i\, \phi\,{\bf\nabla}\cdot{\bf D}\right].  
\end{equation}
The scalar field $\phi$ is conjugate to ${\bf \nabla}\cdot {\bf D}$ and  is
associated with the nodes.  We integrate over the field ${\bf D}$ in eq.
(\ref{Z_phi}) and find an expression in terms of the field $\phi$.  From
here on we assume $\b=1$, or equivalently we absorb the temperature in the
units of energy.
\begin{equation}
\label{Z_phi_only} \mathcal{Z} =
\left(\prod_{links} \e_l^{1/2}\right) \int\mathcal{D}[\phi]\delta(\phi_N)
\,\exp\left[-\frac{1}{2}\sum_{links}\e_l\left({\bf
      \nabla}\phi\right)^2\right]. 
\end{equation} Eq. (\ref{Z_phi_only}) reduces the
problem to that of calculating the partition function of a single scalar field
with $\e$-dependent gradient energy \cite{Dean1}.  The only subtlety is the $\delta$-function
term in the measure.  We will see below that its effect is to remove the
integration over the constant mode, and to contribute an extra overall volume
factor.

Eq.~(\ref{Z_phi_only}) leads to a free energy containing several kinds of
contribution. We wish to understand the lattice dependence of the various
terms and remove those that diverge as we go to the continuum limit and scale
as $(V/s^3)$, where $V$ is the volume and $s$ the lattice spacing.  What
remains after taking care of these divergences corresponds to surface
interactions \cite{Ninham,Dean1}. To separate these contributions we rescale
$\phi$ to $\frac{\phi}{\sqrt{\chi}}$, with $\chi$ an arbitrary, positive
function that, is associated with nodes.  Then, from eq.~(\ref{Z_phi_only})
\begin{equation} 
\label{chi} 
\mathcal{Z} = \prod_{links}\e_l^{1/2}
  \prod_{nodes}\frac{1}{\chi^{1/2}}\;\int\left(\prod_{nodes} d
    \,\phi\right)\delta\left(\frac{\phi_N}
    {\sqrt{\chi_N}}\right)\exp\left[-\frac{1}{2}\sum_{links}\e_l\left({\bf
        \nabla}\frac{\phi}{\sqrt{\chi}}\right)^2\right].  
\end{equation}
We now focus our attention on the integral and on the remaining delta
function.  The first step is to recover simple Gaussian integrals by an
orthogonal change of variable that makes the exponent diagonal. 
We write $\phi_n = \sum
\a_{ni} a_i$, with $n$ running over nodes, $i=0\ldots N-1$, and $\a_{ni}$ are
expansion coefficients.  Then, the integral in eq.~(\ref{chi}) becomes
\begin{equation}
I=\int \left(\prod_{i=0}^{N-1} da_i\right)\, \exp\left[-\sum_i \l_i
  \frac{a_i^2}{2}\right]\; \delta\left(\frac{1}{\sqrt{\chi_N}}\sum_i a_i
  \,\a_{iN}\right) 
\end{equation}
 where the $\l_i$ are the eigenvalues of the operator
\begin{math}
(-\frac{1}{\sqrt{\chi}}\nabla\e\nabla\frac{1}{\sqrt{\chi}})
\end{math}
and $a_i$ are normalized eigenvectors.  The lowest eigenvalue, $\l_0=0$,
corresponds to the field configuration $\phi =N \sqrt{\chi}$, since for this
configuration the gradient in the exponent acts on the constant vector.
The normalized eigenvector is
\begin{math}
{\bf V_0}={\frac{\left[\sqrt{\chi_1}, \sqrt{\chi_2}, \ldots,
    \sqrt{\chi_N}\right]}{\left(\sum_i\chi_i\right)^{1/2}}}
\end{math}.
 Thus, the $\l_0$ expansion coefficient of $\phi_N$ is
\begin{math}
\a_{0N}={\sqrt{\chi_N}}/{\left(\sum_i\chi_i\right)^{1/2}}
\end{math}.

We now integrate over $a_0$
\begin{equation}
I = \left(\sum_{nodes}\chi\right)^{1/2}
\int \left(\prod_{i\neq 0} da_i\right)\,\exp\left[-\l_i    a_i^2\right] %\\
= \left(\sum_{nodes}\chi\right)^{1/2}\mathrm{det}^*\left[-\frac{1}{\sqrt{\chi}}
\nabla\cdot\e\nabla\frac{1}{\sqrt{\chi}}\right]^{-1/2},
\end{equation}
where the $\mathrm{det}^*$ indicates the product over non-zero eigenvalues.
Thus
\begin{equation}
 \label{Z_chi}
 \mathcal{Z}=\left(\prod_{links}\e_l^{1/2}\right)
\left(\prod_{nodes}\frac{1}{\chi^{1/2}}\right)\;
\left(\sum_{nodes}\chi\right)^{1/2}\mathrm{det}^*\left[-\frac{1}{\sqrt{\chi}}
\nabla\cdot\e\nabla\frac{1}{\sqrt{\chi}}\right]^{-1/2}.
\end{equation}
We will now choose $\chi$ to factorize the partition function into extensive
terms and non extensive terms, separating the terms that diverge in the
continuum limit from those that remain finite.

We specialize to the case of interfaces between uniform dielectric
materials.  We discretize as follows.  When a vertex belongs to a slab of
material, all three links departing in the positive direction from that vertex
belong to the same slab.  Thus, for a plane geometry, a slab of thickness
$a=1$  contains one plane of $L^2$ vertices and $3 L^2$ links.  This
definition is not left-to-right symmetric, but it is convenient for periodic
boundary conditions, as a slab of thickness $a$ of $\epsilon_1$ on a
background of $\epsilon_0$ is the same as a slab of thickness $a'=L-a$ of
$\epsilon_0$ on a background of $\epsilon_1$.  We now chose $\chi$ as
$\chi_n=\frac{1}{3}(\e_{nx}+\e_{ny}+\e_{nz})$. For a system where $\e$ is
piece-wise constant this implies choosing $\chi=\e$.

From eq.~(\ref{Z_chi}) we find the free energy, ${\mathcal F}= -\ln \mathcal{Z}$
\begin{equation}
 \label{free_total}
{\mathcal F} = -\frac{1}{2}\sum_{links}\ln \e_l +\frac{1}{2}\sum_{nodes}\ln\e_n -
\frac{1}{2}\ln\left(\sum_{nodes}\e_n\right)+\frac{1}{2}\ln \mathrm{det}^*
\left[-\frac{1}{\sqrt{\e}}\nabla\cdot\e\nabla\frac{1}{\sqrt{\e}}\right]. 
\label{free_lattice}
\end{equation}
The choice $\chi=\e$, leads to a last term in eq.~(\ref{free_total}) which is
homogeneous of degree zero in $\epsilon$. Thus scaling all values of the
dielectric constant by a factor leave this contribution invariant.  
The other contributions contain $\epsilon$
dependent terms which scale as $N$, or $\ln N$, giving divergences as $V$ or
$\ln V$ in the infinite-volume limit.  If we take the continuum limit keeping
the volume finite, but sending the lattice spacing $s$ to zero, these terms
diverge as $1/s^3$ or $\ln s$.  The determinant also has short-distance
divergence in $V$, but this is independent of $\e$ and it is the same for all
systems. It arises from the short-distance behavior of the operator
$-\e^{-1/2}\nabla\cdot\e\nabla\e^{-1/2}$, which in all regions where $\e$ is a
smooth function of ${\bf r}$ is the same as the short-distance behavior of the
operator $-\nabla^2$.  This contribution is the free energy of the vacuum.  It
can be subtracted defining
\begin{math} 
\mathcal{F}_{reg} = \mathcal{F} - \frac{1}{2}\ln \mathrm{det}^*
\left[-\nabla^2\right].
\end{math}
In this paper we will be working with constant total volume, so this term can
be ignored.  We discuss these issues in detail in section~IV.  The surface
free energy is the last term in eq.~(\ref{free_total}); it only contributes
when $\e$ varies at an interface. We therefore define 
\begin{equation}
 \label{free_surf}
\mathcal{F}_{Surf} \equiv \frac{1}{2}\ln 
\left( \mathrm{det}^*\left[-\frac{1}{\sqrt{\e}}
\nabla\cdot\e\nabla\frac{1}{\sqrt{\e}}\right] 
\Bigg/ \mathrm{det}^*\left[-\nabla^2\right]\right).
\end{equation}
This interaction depends {\sl only on the
  ratio} between the different dielectric constants of the uniform 
components of the physical system. For the water/lipid systems we wish to study this ratio is
  $\sim1.7/80\sim 50$, we can than equivalently choose in our numerical codes
  $\e_1=50$ and $\e_0=1$.
Expression (\ref{free_surf}) is free of short-distance divergences in any region where
$\e({\bf r})$ is smooth, since in those regions the short-distance behavior of
the numerator and denominator of eq.~(\ref{free_surf}) are the same.  However,
if $\e$ undergoes a sharp transition at a surface, there is an additional
divergent contribution which can be ascribed to the self-energy of the
interface between the two regions; a sharp interface affects modes of
arbitrarily short wavelengths, therefore the large eigenvalue asymptotics of
the numerator and denominator of eq.~(\ref{free_surf}) do not cancel.  This
divergence is qualitatively different from the volume divergences discussed
earlier in this section. While the latter are the same in any continuum field
theory and are usually dealt with using renormalization theory, the surface
self-energy divergence is related to having defined the field theory on a
\emph{singular background} in which $\e(r)$ is discontinuous
\cite{Jaffe1,Jaffe2}.  In real physical systems treating the dielectric
constant as discontinuous is an oversimplification; any transition between two
material occurs over a small, but finite, distance.  On the lattice, the
lattice spacing gives a natural cut-off and the contributions coming from the
surface self-energy scale as $1/s^2$.  In this work we focus on the interaction
energy between \emph{separate} surfaces, so we will subtract the surface
self-energy by taking free energy differences between systems where the
surfaces are rigidly translated but do not change shape.

Eq.~(\ref{free_lattice}) is the fundamental equation we will use when
extracting the surface interaction from the numerical results.  We first
determine the full partition function directly, either by matrix
diagonalization, or by Monte Carlo simulation. Using eq.~(\ref{free_total}) we
extract the pure surface interaction eq.~(\ref{free_surf}), by subtracting the
extra terms in (\ref{free_total}), which depend on the volume of the system
and/or on the lattice. Explicitly
\begin{equation}
\label{free_surf_comp}
\mathcal{F}_{Surf} = \mathcal{F} + \frac{1}{2}\sum_{links}\ln(\e)-
\frac{1}{2}\sum_{nodes}\ln(\e)+\frac{1}{2}\ln\left(\sum_{nodes}\e\right),
\end{equation}
where for simplicity we do not include the vacuum contribution from
$-\nabla^2$.

\section{Numerical evaluation of Free energies}\label{numerics}

We now compare two methods of evaluating the free energy,
eq.~(\ref{free_total}), of the discretized system.  The first method is a
direct evaluation of the determinant
\begin{math}
\mathrm{det}^*\left[-\frac{1}{\sqrt{\chi}}\nabla\cdot\e\nabla\frac{1}
  {\sqrt{\chi}}\right]^{-1/2}
\end{math}.  The second is an implementation of the Monte Carlo algorithm,
introduced in \cite{Igor}.  We consider a cubic box of side $L$, with periodic
boundary conditions, with a cubic lattice of spacing $s=1$.

For small systems we evaluate the determinant 
\begin{math}
\mathrm{det}^*\left[-\frac{1}{\sqrt{\chi}}\nabla\cdot\e\nabla\frac{1}
{\sqrt{\chi}}\right]^{-1/2}
\end{math}
with standard matrix methods.  If we set $\chi=1$; the partition function
(\ref{Z_chi}) then takes the form
\begin{equation} \label{Z_matrix_0}
\mathcal{Z}=
\left(\prod_{links}\e^{1/2}\right)
L^{3/2}
\,\mathrm{det}^*\left[-\nabla\cdot\e\nabla\right]^{-1/2}
\end{equation}
We write the exponent in (\ref{Z_phi_only}) as a symmetric matrix acting on the
field $\phi$,
\begin{math}
\sum_{nodes}\phi_i M_{ij}\phi_j, 
\end{math} 
where the non-zero elements of $M$ are given by 
\begin{eqnarray}
 M_{i,i} &=& \sum_{nn=1}^6 \e_{i,nn} \nonumber \\
 M_{i,nn} &=& -\e_{i,nn}
 \end{eqnarray}
 where $\e_{i,nn}$ indicates the value of the permittivity on the link
 connecting site $i$ with the nearest neighbor in consideration.  $M$ is a
 symmetric matrix of dimension $L^3\times L^3$.

The determinant is calculated as the product of the eigenvalues of $M$,
$\Lambda_i$. The free energy coming from
$\mathrm{det}^*\left[-\nabla\cdot\e\nabla\right]$  is
\begin{equation} \label{F_det}
\mathcal{F}_{det} = +\frac{1}{2}\sum_{i=1}^{L^3-1} \ln\Lambda_i,
\end{equation}
where now the sum extends only from $i=1$ to exclude the eigenvalue $\Lambda_0
= 0$.  The free energy, from (\ref{Z_matrix_0}) is then
\begin{equation} \label{free_M}
\mathcal{F} = \mathcal{F}_{det} -\frac{1}{2}\left[\sum_{links}\ln \e- \ln(L^3)\right].
\end{equation}
Comparing eq.(\ref{free_M}) and(\ref{free_surf_comp}), we see that the surface
free energy is
\begin{equation}
\label{free_M_2}
\mathcal{F}_{surf} = \mathcal{F}_{det} - \frac{1}{2}\sum_{nodes}\ln  \e+
\frac{1}{2}\ln\left(\sum_{nodes}\e\right) - \frac{1}{2}\ln(L^3).
\end{equation}

The last thing we need to take care is the constant, $\e$-independent
contribution $\ln\mathrm{det}^*(-\nabla^2)$, which was discussed in
section~\ref{sec_lattice}, that represents the vacuum contribution (see also
the Appendix for details).  We can subtract this term by hand by computing it
explicitly, or equivalently always consider free energy differences.  We adopt
the second strategy, so that the constant factor $\ln L^3$ entering in the
last term of eq.~(\ref{free_M_2}) can also be ignored.

As an alternative procedure,
we have used the method of \cite{Igor} to measure the free energy using
thermodynamic integration and Monte Carlo simulation.  We sample the partition
function (\ref{Z_deltaD}) using a collective worm algorithm \cite{MaggsW} to
update the field ${\bf D}$. We obtain free energy differences between two
system: the one under study, and a reference system.  In order to avoid surface 
divergences, the reference system has to
be chosen as a rigid translation of the original system.  In the example of
parallel planar slabs, that we are going to analyze in detail, we take as
``reference'' the system where the surfaces are at maximum distance (i.e.
$L/2$, in periodic boundary conditions).

The free energy obtained by the simulation, $\mathcal{F}_{sim}$, gives
directly an evaluation of the left hand side of eq.  (\ref{free_total}).
Using eq. (\ref{free_surf_comp}), we conclude that the surface contribution is
\begin{equation}
\Delta\mathcal{F}_{surf} = \Delta\mathcal{F}_{Sim} - 
\Delta\left[-\frac{1}{2}\sum_{links}\ln(\e)+\frac{1}{2}\sum_{nodes}\ln(\e)-
\frac{1}{2}\ln\left(\sum_{nodes}\e\right)\right].
\end{equation}

\section{Analytic results} 
\label{analytics} 

\begin{figure}[t]  
\begin{center}
\includegraphics[width=8cm]{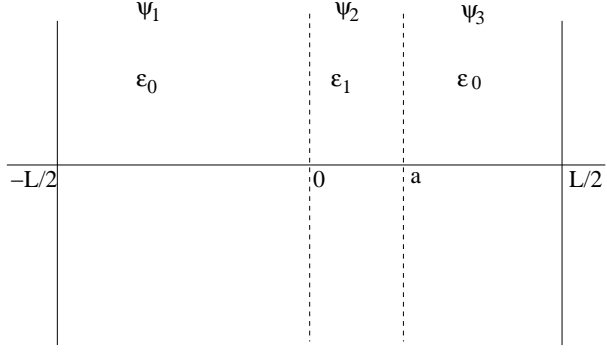} 
\caption{{\small One slab geometry.}} \label{fig_1}
\end{center}
\end{figure}

The above discretization methods can be applied in arbitrary geometry. To
calibrate them we will apply them to a flat slab, and compare the
results with an analytic expression for the surface free energy,
eq.~(\ref{free_surf}), in the continuum limit.  We consider a single-slab,
piece-wise uniform system, periodic along the direction perpendicular to the
slab, and infinite in the two transverse directions. Its geometry is shown in
figure~\ref{fig_1}. Analytical results are well known for the similar system
which is infinite in all three directions \cite{Kampen,Ninham,Dean1}, but, to
our knowledge, the result we derive here for the periodic system is new.

Starting from the constrained partition function on the continuum,
eq.~(\ref{Z}), we derive the free energy of the periodic system following a procedure
similar to \cite{Kampen,Ninham}.  However, our derivation does not extract the
classical result as a limit of a quantum system; it considers the classical
system from the start.  We believe that in such treatment the physics is more
transparent and we are able to better control issues such as subtraction of
divergences which, as we have seen, are important for the comparison of the
analytical results with numerical data.  In particular, since the final
expression we derive is free of divergences, but the starting point, eq.
(\ref{Z}), is not, we have to make sure that the finite quantities we
calculate are consistent with the corresponding finite quantities defined on
the lattice and computed through the numerical methods.  We will show that
expression (\ref{free_surf}) is essentially finite, (i.e.\ up to the
--unphysical-- divergent surface self-energy that appears in singular
backgrounds, that will be discussed separately), so it is meaningful to compare
it with numerical results.

We will compute the free energy per unit transverse area. This quantity
stays finite in the limit of infinite transverse size, and   
 is defined, following eq.~(\ref{free_surf})
\begin{eqnarray}
 \label{free_Tr}
\mathcal{F}_{Surf} &=&  \frac{1}{2\,L_t^2}\ln {\mathrm{det}^*
\left[-\e^{-1/2}\nabla\e\nabla\e^{-1/2}\right]\over \mathrm{det}^*\left[-\nabla^2\right]}\\
&=& \frac{1}{2\,L_t^2}\left\{\mathrm{Tr}^*\ln
\left[-\e^{-1/2}\nabla\e\nabla\e^{-1/2}\right] -
\mathrm{Tr}^*\ln\left[-\nabla^2
\right]\right\}, \label{Trace}\\
&\equiv& \tilde{\mathcal{F}} - \mathcal{F}_0
\end{eqnarray}
where $L_t$ is the linear dimension of the system along the uniform directions, 
and the limit $L_t\to \infty$ is understood.   

Determinants are cyclic invariant so that $\det(-\e^{-1}\nabla\e\nabla)$ is equivalent
to $\det(-\e^{-1/2}\nabla\e\nabla\e^{-1/2})$.  In our test case, we find it
more convenient to evaluate
\begin{equation}
 \label{Ftilde}
\tilde{\mathcal{F}} =  \frac{1}{2\,L_t^2}\mathrm{Tr}^*\ln (-\e^{-1}\nabla\e\nabla).
\end{equation}
Let us consider the geometry in fig.(\ref{fig_1}), with $\e({\bf r})$
piece-wise constant.  We take our system to be periodic on $[-\frac{L}{2},
\frac{L}{2}]$.  The surface of the slabs are perpendicular to the $z$ axes. We
separate the eigenvalue equation writing $\psi(x,y,z)=e^{i\r_x x + i \r_y
  y}\psi_z(z)$, with:
\begin{equation}
  \label{eigenvalues_2}
\e^{-1}(\e \r^2 -  \partial_z \e \partial_z)\,\psi_{z} = \L\,\psi_{z},
\end{equation}
and $\r^2=\r_x^2 + \r_y^2$.  \\
The eigenfunction along $z$ is taken separately
in the three regions
\begin{equation}
\psi_\alpha(z) = A_\alpha e^{i p z} + B_\alpha e^{-i p z} \label{psi1} \\
\end{equation}
with $\alpha=1,2,3$.
Integrating the
  eigenvalue equation over a small interval we find that $\psi$ satisfies the same
  boundary conditions as scalar electrostatic potential $\phi$:
\begin{eqnarray}
\psi_-(z) &=& \psi_+(z), \label{psi4}\\
\e_-\partial\psi_-(z) &=& \e_+\partial\psi_+(z), \label{psi5}
\end{eqnarray}
which are derived by integrating (\ref{eigenvalues_2}) across the boundaries.
Here, $+$, and $-$ refer to left and right side of the interface.

Eq.(\ref{eigenvalues_2}) gives
\begin{math}
\r^2 + p^2 = \L \label{p_and_L}
\end{math} and
\begin{math}
\r^2 + q^2 = \L \label{q_and_L},
\end{math}
from which we immediately deduce $p=q$.  Moreover, we notice that $\L$ has to
be real and positive.  Indeed, consider an $\e(r)$ everywhere positive, with
the eigenvalues equation
\begin{math}
-(\nabla\e\nabla)\psi = \e\L\psi
\end{math}.
Then, 
\begin{math}
0\leq\int\e|\nabla\psi|^2 =
-\int\psi^*(\nabla\e\nabla)\psi-\int\nabla\left[\psi^*(\e\nabla\psi)\right]
\end{math}.  The last integral is zero because of the matching condition
(\ref{psi5}), leaving us with:
\begin{math}
0\leq - \int\psi^*(\nabla\e\nabla)\psi = \Lambda \int\e|\psi|^2
\end{math}.  As a consequence $q$ has to be either real or purely imaginary,
and in the latter case $|q| \leq \r$.

Inserting the functions (\ref{psi1}) in the eigenvalue equation
(\ref{eigenvalues_2}) and using the boundary conditions, we find
\begin{equation}
 \label{eigen_eq}
\left(\frac{\e_1-\e_0}{\e_1+\e_0}\right)^2\frac{1-\cos(q(L-2a))}{1-\cos(qL)}=1.
\end{equation}
This equation determines the eigenvalues $q_n$ of the effective
one-dimensional eigenvalue problem, eq. (\ref{eigenvalues_2}).  The surface
contribution to the free energy per unit area from~(\ref{Ftilde})
is\footnote{The replacement $L_t^2\to (2\pi)^{-2}$ is understood by writing
  down the trace as a sum of diagonal terms on the eigenfunction $\psi_{q,p}$
  with the operators depending on the two transverse momenta $\r$ and $\r'$,
  and integrating over $\r'$.  We find an infrared divergence of the form
  $\delta^2(0)$ which can be regularized introducing a cut-off in the
  positions. This gives a multiplicative factor $(L_t/2\pi)2$.}:
\begin{eqnarray}
\tilde{\mathcal{F}} &=& \frac{1}{2\,L_t^2}
\mathrm{Tr}\ln(-\e^{-1}(\e \nabla_t2 +  
\partial_z \e \partial_z)) \nonumber \\
&=&  \frac{1}{8\pi^2}\int_{0}^{\inf} d^2\r \,{1\over 2}\sum_n \ln(\r^2+q_n^2) \label{Tr_ln},
\end{eqnarray}
where $q_n$'s are the solutions of eq. (\ref{eigen_eq}).  The extra factor 2
in the integrand of the last expression comes from the consideration that the
transformation $q\to -q$ leaves both eq. (\ref{eigen_eq}), and the
wavefunction (\ref{psi1}), invariant up to a renaming of the
coefficients.  Summing over all solutions of (\ref{eigen_eq}) requires
dividing by 2 to avoid double-counting. An equation similar to (\ref{Tr_ln})
holds for $\mathcal{F}_0$ that we need to subtract, with the $q_n$'s replaced
by the appropriate eigenvalues for the uniform system.

\begin{figure}[t] 
\begin{center}
\includegraphics[width=6cm]{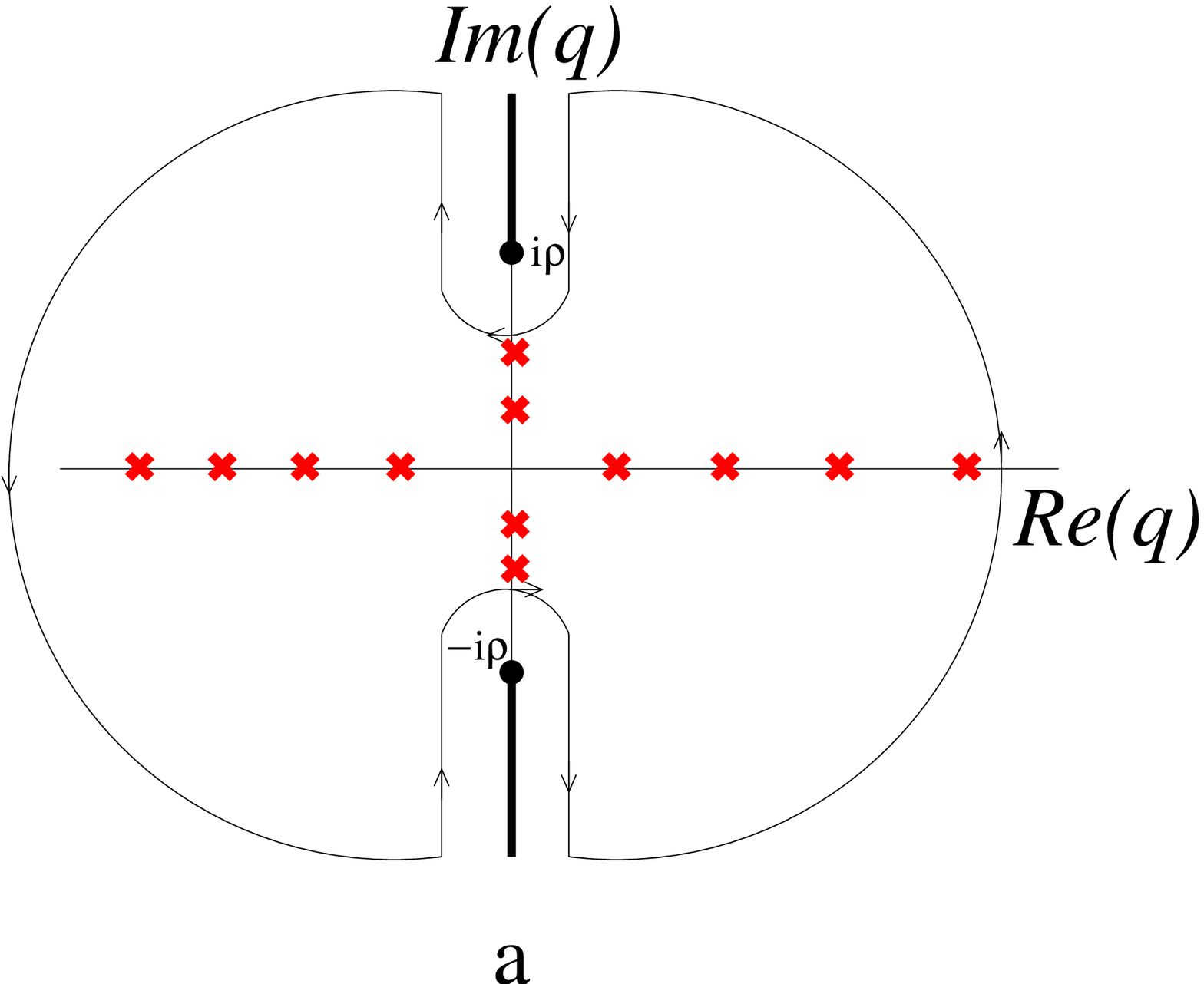} \hspace{1cm}
\includegraphics[width=6cm]{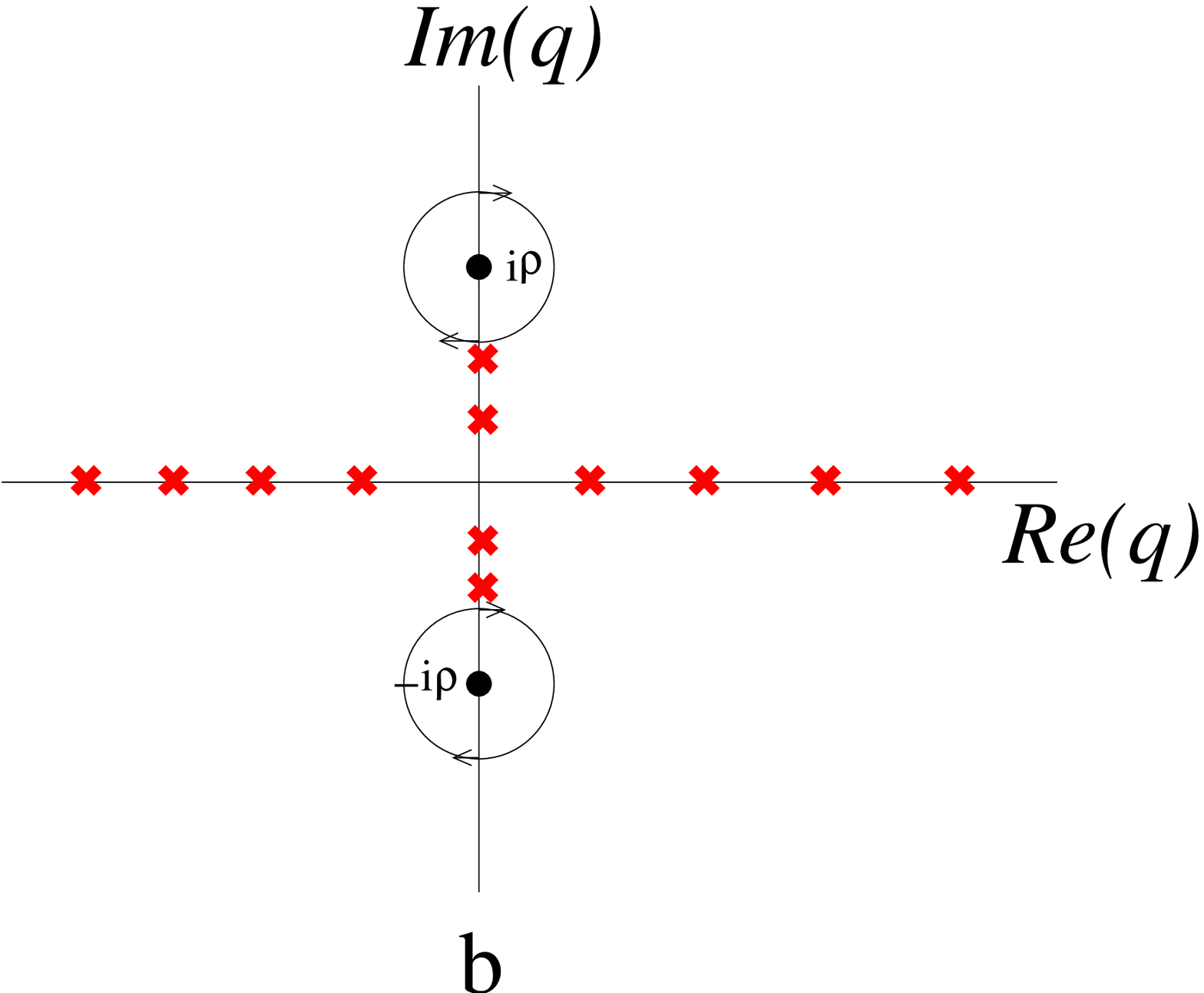}
\caption{{\small (a) Contour of integration.  The zeros of $Q(q)$ are the big X. 
(b) Contour after integration by parts.}} \label{fig_2}
\end{center}
\end{figure}

In the Appendix we show that the free energy, eq.~(\ref{Trace}), can be
written in terms of a ``spectral function'' on the complex $q$-plane, $Q(q)$:
\begin{equation}
\label{contour}
\mathcal{F}_{surf} = \tilde{\mathcal{F}} - \mathcal{F}_0 = 
\frac{1}{16\pi^2}\int_0^{\inf} d^2\r \oint_\gamma\frac{dk}{2\pi i} \ln(\r^2+k^2)\frac{Q'(k)}{Q(k)},
\end{equation}
where the integration contour $\gamma$ encloses all the complex solutions of
the eigenvalue equation (\ref{eigen_eq}), for example the path shown in
figure~\ref{fig_2} (a), chosen in such a way that it avoids the branch cuts
$(-i\infty,-i\r)$ and $(i\r,+i\infty)$ from the logarithm.  For the
single-slab configuration, the appropriate function is:
\begin{equation} \label{Q}
Q(q) =1- \left(\frac{\e_1-\e_0}{\e_1+\e_0}\right)^2\frac{1-\cos(q(L-2a))}{1-\cos(qL)}.
\end{equation}
As shown in the Appendix, this choice corresponds to subtracting the energy of
uniform empty space, the second term in eq. (\ref{Trace}).

To arrive at an explicit expression for  eq.~(\ref{contour}), we first rewrite 
the argument of the $\rho$-integral, 
\begin{equation}
f(\rho) = \oint_\gamma\frac{dk}{2\pi i} \ln(\r^2+k^2)\frac{Q'(k)}{Q(k)}. 
\end{equation}
 Rather than evaluating directly $f(\rho)$, let us calculate its derivative: 
\begin{equation}
\label{cont_int_der}
{d\over d \r}f(\r) = \oint_\gamma \frac{dk}{2\pi i}
{2\r \over \r^2+k^2}\frac{Q'(k)}{Q(k)} = \oint_\gamma\frac{dk}{2\pi i}
{2\r \over \r^2+k^2}{d\over d k}\ln Q(k),
\end{equation}
 The integral along  the large circle at infinity vanishes, and the only
 non-zero contribution  comes from the paths along the imaginary axis.  
We  deform the contour to two circles enclosing the
two poles in $\pm i\r$, as shown in figure~\ref{fig_2} (b). 

Eq.~(\ref{cont_int_der}) is evaluated using the residue theorem:
\begin{equation}
\label{fbar}
{d f \over d\r} = {d\over d\r}\ln Q(i\r)+ {d\over d\r}\ln Q(-i\r), 
\end{equation}
whence 
\begin{equation}
\label{frho}
f(\r) = 2\ln Q(i\r) + 2\bar{f}
\end{equation}
where $\bar{f}$ is an integration constant, and we have used $Q(q) = Q(-q)$.  
Inserting eq.~(\ref{frho}) in  eq.~(\ref{Trace}), we  obtain 
\begin{equation}
 \label{F_general_div}
\mathcal{F}_{Surf}=\frac{1}{4\pi} \int_0^{\inf}\r \ln Q(i\r)\,d\r +
\frac{\bar{f}}{4\pi}
 \int_0^{\inf}\r \,d \r 
\end{equation}
The last term is still quadratically divergent. If we introduce a cut-off
$\omega$ in momentum space, the last integral is $O(\bar{f}\omega^2)$, the
same scaling as a short-distance surface divergence.  This is the explicit
manifestation of the divergent surface self-energy described in
section~\ref{numerics}, and it is caused by the singular background,
characterized by a sharp transition in the dielectric profile.  This
divergence can be subtracted by setting the integration constant $\bar{f}=0$,
as it does not depend on the dielectric constants.  Notice however that the
divergence will depend, in general, on the shape of the surfaces considered,
so if we want to compare the free energies of systems whose interfaces have
different geometries, we must treat this divergence more carefully.  The
self-energy can be interpreted as the \emph{finite tension} of the surface
interface, and one has to take it into account if one wishes to treat the
surfaces \emph{dynamically}.
    
Setting the last term in eq.~(\ref{F_general_div}) to zero we find
\begin{equation}
 \label{F_general}
\mathcal{F}_{Surf}=\frac{1}{4\pi} \int_0^{\inf}\r \ln Q(i\r)\,d\r 
\end{equation}
Eq.(\ref{F_general}) is gives us the free energy once we know the eigenvalue
equation.  In the appendix we show how the choices we made uniquely determine the
function $Q(q)$ as its zeros correspond to the solutions of the eigenvalue
equation, and its poles correspond to the eigenvalues of the uniform system.
These two requirements are equivalent to asking that the subtraction we
performed to arrive at eq. (\ref{F_general}) is exactly the second term in
eq.~(\ref{Trace}).

Using eq. (\ref{Q}), the surface free energy becomes
\begin{equation}
 \label{free_Ch}
\mathcal{F}_{Surf}=\frac{1}{4\pi} \int_0^{\inf} \r
\ln\left[1-\left(\frac{\e_1-\e_0}{\e_1+\e_0}\right)^2\;
\frac{1-\mathrm{Ch}[\r(2a-L)]}{1-\mathrm{Ch}[\r L]}
\right]\;d\r.
\end{equation}
The integral over $\r$ is finite for separations $a>0$; the integrand varies
as $\r e^{-2a \r}$ for large $\r$. Thus the surface free energy,
eq.~(\ref{free_surf}), is free of short-distance divergences and can be used
in numerical computations.

In the limit $L \to \inf$, eq. (\ref{free_Ch}) reduces to  \cite{Ninham2,Dean1}
\begin{math}
 \label{free_inf}
\mathcal{F}_{Surf}=\frac{1}{4\pi}\int_0^{\inf} \r
\ln\left[1-\left(\frac{\e_1-\e_0}{\e_1+\e_0}\right)^2 e^{-2\r a}\right] d\rho.
\end{math}
The method of this section and the Appendix can be extended to systems
composed of an arbitrary number of slabs through the use of transfer matrices
\cite{Ninham2}.

\section{Numerical Results}\label{results}

We used both matrix diagonalization and Monte Carlo simulation to study the
slab geometry of section~\ref{analytics} in order to calibrate the numerical
methods and evaluate discretization errors.  We firstly consider a volume
$V=N=12^3$ with lattice spacing $s=1$.  The dielectric constant of the slab is
$\e_1=2$ with background dielectric constant $\e_0=1$.  These values of
  $\e$ {\sl do not correspond to any interesting physical system}, and are far from
  the 1:50 ratio we seek when studying water/lipid systems. However one of the
  methods with which we wish to compare (Monte Carlo simulation) becomes
  inefficient for large dielectric contrasts. It is thus more convenient to
  work with reduced contrast in order to compare results.  

\begin{figure}[t] 
\begin{center}
\includegraphics[width=10cm]{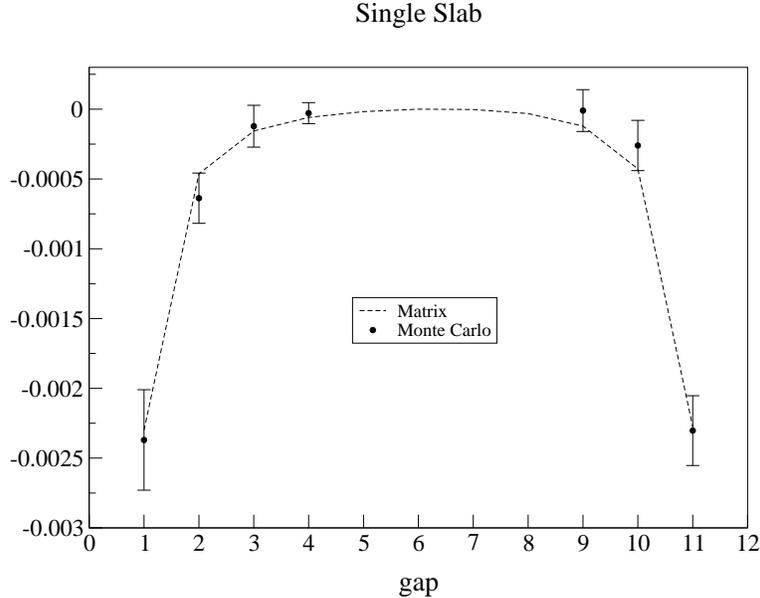} 
\caption{\small $\Delta\mathcal{F}_{Surf}$ as a function of surface separation
  from matrix diagonalization (solid) and Monte Carlo (points).  The system
  size $L=12$, dielectric constants $\e_0=1$ and $\e_1=2$.  The reference
  system is taken to be a system with maximal slab thickness, i.e.  $a' =
  L/2$.} \label{fig_3}
\end{center}
\end{figure}

The matrix diagonalization is performed using the built in methods of Matlab.
It requires about one minute on a 2GHz workstation to calculate the free
energy as a function of the gap $G$, with $G = 1,\ldots,11$.  The Monte Carlo
simulation was run with 40 points for the thermodynamic integration.  For each
gap measuring the free energy requires about one hour of simulation.
Figure~\ref{fig_3} shows that the agreement between the two methods is within
error bars.

The difference in computer resources needed in these two calculations of the
free energy shows that Monte Carlo is not the best method for
studying the large distance tails of the interaction between dielectric media.
The surface interactions we want to measure are three to four orders of
magnitude smaller than the self-energy; their extraction requires extremely
precise measures which are hard to achieve with statistical evaluations. We
can estimate the scaling of the resources needed in Monte Carlo as follows:
The interaction between two interfaces scales as $L^2/G^2=O(1)$ which is to be
compared to the volume free energy which scales as $O(V)$. Thus we require
$O(V^2)$ Monte Carlo sweeps of cost $O(V)$ to generate statistically useful
results for a single integration point. In addition, as the system size grows,
more simulations points are needed for the thermodynamic integration.
Considering all these factors, we see that the Monte Carlo method
requires an effort that scales worse than $L^9$.

Matrix diagonalization is not affected by statistical noise.  The complexity
is dominated by the evaluation of the determinant: $O(V^3)=O(L^9)$ with dense
matrix routines.  The real constraint for matrix diagonalization turns out to
be the memory required for holding dense matrices.  With 1GB memory avaliable for
computation, using standard routines, the largest system size we can consider
is $L_{max}\sim 25$; memory usage scales as $O(V^2)=O(L^6)$.  In order to
study larger, more interesting systems we now specialize to objects which are
translationally invariant in one direction. Because of the symmetry one can
then use Fourier analysis to simplify the numerical problem. When we do this
we find that we need only find the eigenvalues of a matrix of dimensions
$L^2\times L^2$ instead of $L^3 \times L^3$.

We write $\phi=\sum_{q_z} \phi_z^{(q_z)}\phi_{xy}^{(q_z)}$, where
$\phi_z^{(q_z)}$ are plane waves with $q_z = 2\pi n/L$ and $n = 0,1,\ldots,L$,
corresponding to the eigenvalue $2(1-\cos(q_z))$ on a periodic lattice. Using
this form for the eigenfunctions we find
\begin{equation}
\phi\, M\, \phi = \sum_{q_z}\phi_{xy}^{(q_z)}\,\tilde{M}(q_z)\,\phi_{xy}^{(q_z)}
\end{equation}
where the non-zero elements of the reduced matrix $\tilde{M}(q_z)$ take the form
\begin{eqnarray} \label{matrix_FT}
 \tilde{M}_{i,i}(q_z) &=& \sum_{nn=1}^4\e_{i,nn}+2\e_z(1-\cos(q_z))\nonumber \\
\tilde{M}_{i,nn}(q_z) &=& -\e_{i,nn}
\end{eqnarray}
and $\ln\det^*M = \sum_{q_z}\ln\det^*\tilde{M}(q_z)$.  The sum over nearest
neighbors runs only along the $x$ and $y$ axes, $\e_z$ indicates the value
of $\e$ along the uniform direction $z$.  Using this approach, the new limit
in size due to memory constraints becomes $\tilde{L}_{max} = L_{max}^{3/2}
\sim 125$.  However, in this case, computing time becomes the ultimate limiting
factor because the determinant evaluation has to be performed $L$ times, once
for each $q_z$, requiring an effort $O(L^7)$.  A system size of about $L=50$
is the largest system we can consider on our same 2GHz processor, which leads
to evaluating the determinant in about one hour.

\begin{figure}[t] 
\begin{center}
\includegraphics[width=10cm]{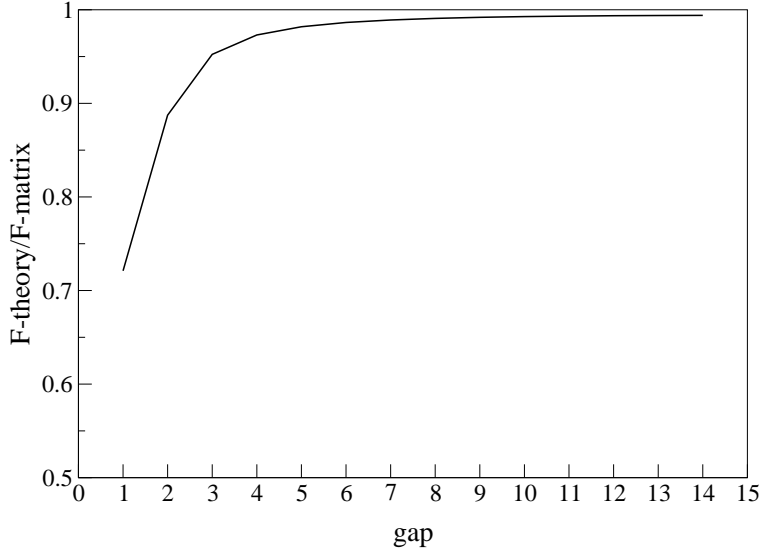} 
\caption{\small Ratio of $\mathcal{F}_{th}$ computed from the analytic formula
  eq. (\ref{free_Ch}), and $\mathcal{F}_m$ calculated from the matrix. The
  transverse dimensions are $L_x = L_y = 60$, while the longitudinal direction is
  $L_z=30$. } \label{fig_mthratio}
\end{center}
\end{figure}

We used this technique to compare the result obtained with matrix methods to
our analytical calculations. Our analytic result eq.(\ref{free_Ch}) is valid in the limit
that the transverse directions of the system are large compared to the
longitudinal direction. We thus study a system of dimensions $60 \times 60
\times 30$, where the smaller number refers to the longitudinal direction, and
compare it to eq. (\ref{free_Ch}) for $\e_1=50$ and $\e_0=1$.  Results are
shown in figure \ref{fig_mthratio} where we have plotted the ratio of the
analytical result over numerical result as a function of surface separation.
We see that the agreement is rather poor for the smallest separations, as
might expected in a discretized system. For larger gaps the agreement
significantly improves

We now use the Fourier-matrix method to study the interaction between parallel
grooves in dielectric surfaces. Using microfabrication techniques such grooves are
easy to manufacture; recent theoretical studies have investigated the effects
of corrugations on quantum Casimir free energies \cite{Emig1,Emig2}.  We also
note that one can study the Casimir energy as a function of lateral phase
shift between the groove corrugations.  Experimental evidence for lateral
force was presented in \cite{Chen}.

For sinusoidal grooves analytic results were found using a second order
perturbation in the amplitude parameter while for rectangular
grooves an exact numerical solution was found.  The effect of corrugation
wavelength and amplitude in relation to surface separation was investigated
and it was found that for corrugation lengths much bigger than gap $G$,
Proximity Force Approximation (PFA) works well while in the limit of
corrugation lengths much smaller than the amplitude a small distance regime is
reached \cite{Emig3}.  The first correction term to the parallel plate free
energy, $\delta = \mathcal{F}/\mathcal{F}_{flat} - 1$, was determined.  It was
found that in the limit of the corrugation length $\l\gg G$ the correction
$\delta$ to the Casimir energy depends only on the corrugation amplitude, and
it is proportional to $G^{-1}$.  In the opposite situation of $\l\ll G$ the
correction to the energy depends on both wavelength and amplitude and $\delta
\sim G^{-2}$.

Our numerical approach is valid for any groove profile, or indeed with any
$2+1$ dimensional interfacial profile, and for all separation regimes. We
first consider the interaction of rectangular groove and a planar interface.
We analyzed the dependence of the free energy on the corrugation wavelength
$\l$ and on the surface separation $G$.  In accordance to the analytical
results for quantum Casimir, in the limit $\l\gg G$ we find the Proximity
Force Approximation gives results in agreement with our numerical results.
According to this widely used approximation, one considers the interaction of
a small portion of the surface with a corresponding portion on the other
surface which sits vertically above, assuming the surface is locally flat.
The interaction between the two ``plaquettes'' is taken as the interaction of
two parallel surfaces at the same vertical distance, and the contribution from
all plaquettes is then added to obtain the total interaction.  In a system of
$V=40^3$, for the particular case of groove depth $H=4$, and gap $G=6$, at maximal wavelength
$\l=20$ the agreement between the PFA prediction and the direct numerical
result is around $90\%$.  As $\l$ becomes smaller and comparable to the gap,
the agreement drops to less than $20\%$.  We have also investigated the free
energy behavior in the limit $\l\ll G$.  We have indications that, as in the
quantum system, the correction term $\delta$ goes as $1/G$, but in order to be
more certain one would need to study a larger system where ratios $G/\l
\sim 10^2$ can be reached.
\begin{figure}[t] 
\begin{center}
\includegraphics[width=8cm]{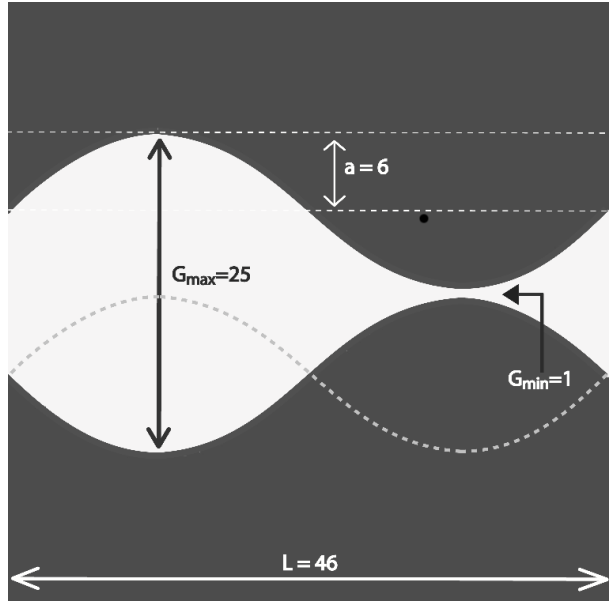} 
\caption{\small Large sinusoidal grooves at $S=L/2$. The dashed profile indicates
the position of the lower surface in the configuration $S=0$.} \label{sin_groove}
\end{center}
\end{figure}

\begin{figure}[t] 
\begin{center}
\includegraphics[width=10cm]{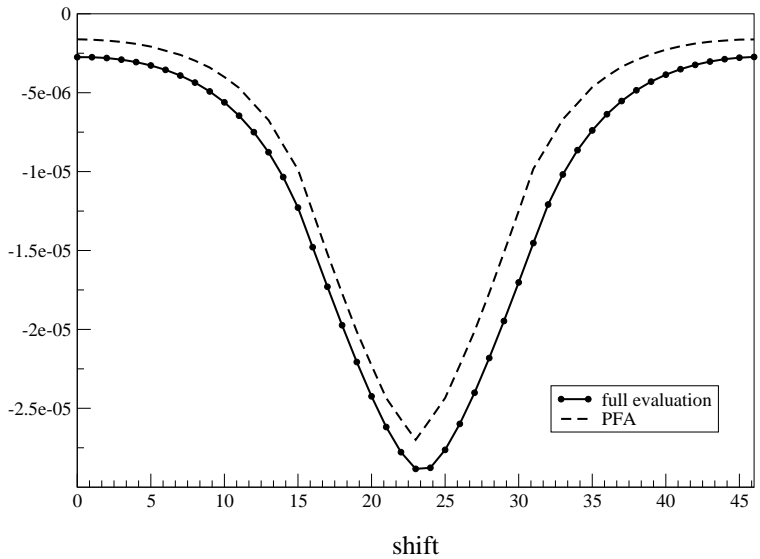} 
\caption{\small Free energy as a function of phase shift.} \label{free_sh}
\end{center}
\end{figure}

We now consider a sinusoidal groove, figure~\ref{sin_groove}, with system size
$L=46$, amplitude of oscillation $a=6$, wavelength $L$.  The minimal
distance $G_{min}$ between the grooves is reached when their relative phase is
$\pi$, and is equal to $1$.  
In this configuration the vertical distance between the surfaces varies from
$G_{min}=1$ to $G_{max}=25$. 
The shift $S$ is measured in units of the lattice
spacing as the displacement from the position where the grooves are perfectly
aligned.  
Letting the top groove shift laterally over the bottom groove 
from the $S=L/2$ configuration (shown in figure~\ref{sin_groove}) we measure the free energy 
as a function of the shift,$\mathcal{F}(S)$.  
$\mathcal{F}(S)$ gives information about lateral Casimir
forces.  When looking at the effect of the shift one expects the free energy
to vary because the distribution of vertical distances changes as one
groove is laterally displaced. Indeed when the shift is zero all the points of
the opposing grooves have the same distance $2a + G_{min}$, while for $S=L/2$,
the distance between any two vertically aligned points varies from $G_{min}$
to $G_{max}=G_{min}+4a$.  But beside this projection of the vertical Casimir
interaction along the direction of the displacement, one expects to also detect lateral
interactions between curved surfaces.  Due to the non-additivity of Casimir
forces, and because of the large deformation of the groove with respect to the
flat geometry, these two contributions can not be disentangled.  In figure
\ref{free_sh} we present the results obtained with the reduced matrix
diagonalization method, that gives the global interaction, and compare it to
PFA, that only considers the effect of the \emph{local} vertical displacement.
PFA underestimates the free energy by factors that vary from around $10\%$
when the grooves are mirroring (phase shift $\pi$), up to $40\%$ for phase
shifts close to zero.  This is what one would expect since PFA misses to
consider collective effects, which are more important when larger positions of
the surfaces are close.
These violations of the proximity-force-approximation are not peculiar to the 
classical regime and similar effects are also found for the quantum system for
significantly non-flat geometries \cite{Emig1,Gies}.

\section{Conclusions} 
\label{conclusions} 

We have presented a formulation of fluctuation induced interactions between
dielectrics discretized to a lattice.  We analyzed how the contributions to
the on-lattice partition function depend on the discretization and on the
volume, and we have derived an appropriate definition for the surface
interaction, which stays finite in the continuum limit.  We compared two
numerical methods to compute the free energy of surface interactions.  The
first is a direct evaluation of the matrix determinant.  The second is a Monte
Carlo simulation together with thermodynamic integration.

While the constrained partition function allows the simulation of systems
including full long-ranged Casimir interaction it turns out to be a rather
inefficient method for extracting the asymptotic interactions between bodies.
The non-extensive surface interactions are rather easily lost in statistical
noise. We consider that the use of matrix methods has considerable promise.
Already interesting problems can be studied in the $2+1$ dimensional in of
translationly invariant systems, such as grooves, or blades. We note that we
have only used the simplest dense matrix methods. We anticipate that the use
of more specialized sparse matrix solvers should allow the study of larger
physical systems, without the requirement of translation invariance in one
dimension.

Generalizations including the study of more elaborate dielectric functions,
that include scale dependent dielectric response \cite{Tony-Ralph} as well as
quantum effects will be presented in future papers.

\section*{Acknowledgments}

We would like to thank R. L. Jaffe and R. Golestanian for comments and
discussion.  This work was partially supported by Volkswagenstiftung, by ANR
grant NT05-1-41861, INTAS grant, 03-51-6346, RTN contracts MRTN-CT-2004-005104
and MRTN-CT-2004-503369, CNRS PICS 2530 and 3059 and by an European Excellence
Grant, MEXT-CT-2003-509661.  F. N. is supported by an European Commission
Marie Curie Intra European Fellowship, contract MEIF-CT-2006-039369

\appendix
%\section*{Appendix}
\section{Regularization of  Ultra violet divergences}
\renewcommand{\theequation}{A.\arabic{equation}} \setcounter{equation}{0}

In this Appendix we analyze the short-distance divergences in the parallel
plate geometry discussed in section IV. We address the subtraction of
divergent vacuum energy and the divergent surface self-energy, showing that
the latter arises from the singular nature of the background.
 
%\subsection{Regularization of the Vacuum Energy}
%\setcounter{\theequation}{00}
In the continuum limit, there is always an $\e$-independent divergent
contribution in the free energy which arises from the existence of modes of
$\phi$ with arbitrarily large-momentum.  To see this, consider a system where
$\e$ is everywhere uniform.  The free energy corresponding to
$\mathrm{det}^*(-{\bf \nabla^2})$ diverges as 
\begin{equation}
 \mathcal{F}\sim{1\over
  2}\sum_n\log \l_n, \qquad \l_n \sim n^2.  
\end{equation}
 This divergence corresponds to large-momentum modes; it is present in the
 single-slab system analyzed in section~\ref{analytics}, as the
 eigenvalues in eq.~(\ref{eigen_eq}) also grow indefinitely.

Consider now the free energy before the subtraction, eq. (\ref{Tr_ln}) 
\begin{eqnarray}
  \tilde{\mathcal{F}} &=& \frac{1}{2\,L_t^2}\mathrm{Tr}^*\ln
  (-\e^{-1}\nabla\e\nabla) \\ 
  &=& \frac{1}{8\pi2}\int_0^{\inf} d^2\r \,{1\over 2}\sum_n \ln(\r^2+q_n^2). 
\label{Tr_ln_app}
\end{eqnarray}
We now write the summation as a contour integral in the complex plane using
Cauchy's theorem,
\begin{equation}
\label{cauchy} \sum_n \ln(\r^2+q_n^2)=
\oint_\gamma\frac{dk}{2\pi i} \ln(\r^2+k^2)\frac{\tilde{Q}'(k)}{\tilde{Q}(k)},
\end{equation}
where the spectral function $\tilde{Q}(q)$ has simple zeros\footnote{In case
  of zeros of order $p$ , the the r.h.s.\ must be multiplied by 1/p} for
$q=q_n$, and the integral is over an appropriate contour $\gamma$ that
encloses all zeros of $\tilde{Q}(q)$.  We have used the fact that whenever
$\tilde{Q}(q)$ has a simple zero, $\tilde{Q}'/\tilde{Q}$ has a simple pole
with unit residue.  Eq.  (\ref{cauchy}) is clearly ill-defined, since the
argument of the summation becomes arbitrarily large.

Notice that, if  $\tilde{Q}(q)$ has a pole of order $p$ at $q=\bar{q}$, 
eq.~(\ref{cauchy}) is invalid, 
since close to $\bar{q}$ we have
\begin{equation}
\label{poles}
{\tilde{Q}'(q)\over \tilde{Q}(q)} \sim -{p\over (q-\bar{q})} 
\end{equation}
and we get an extra contribution to the r.h.s.  So, even formally, expression
(\ref{cauchy}) holds only if the spectral function has zeros corresponding to
the eigenvalues of the problem at hand, and no other zeros {\em nor poles}.
However we can see from eq. (\ref{poles}) that introducing simple poles
$\bar{q}_n$ in the spectral function is equivalent to {\em subtracting}\/ the
free energy $\bar{\mathcal{F}}$ of a system that has eigenvalues $\bar{q}_n$.
In this case therefore we formally have 
\begin{equation}
\label{difference} \sum_n
\ln(\r^2+q_n^2) - \sum_m \ln(\r^2+\bar{q}_m^2) = \oint_\gamma\frac{dk}{2\pi i}
\ln(\r^2+k^2)\frac{Q'(k)}{Q(k)}, 
\end{equation}
where now $Q(q)$ has $q_n$ as zeros and $\bar{q}_m$ as poles. Depending on the
properties of the spectral function, it can happen that the two terms on the
l.h.s.\ are infinite, but the integral on the r.h.s.\ is finite, so that the
expression (\ref{difference}), corresponding to the difference in free
energies, is meaningful.

Thus, the free energy difference between a  given  system and the vacuum can 
be written using eq. (\ref{difference}), in which 
\begin{equation}
  Q(q) = {\tilde{Q}(q)\over Q_0(q)}, 
\end{equation}
where the zeros of $\tilde{Q}(q)$ and of $Q_0(q)$ are the solutions of the
eigenvalue equations of the system under consideration and of the vacuum,
respectively, and neither function has other zeros or poles. Using Eqs.
(\ref{Tr_ln_app}) and (\ref{difference}) we arrive at 
\begin{equation}
\label{contour_app}
\tilde{\mathcal{F}} - \mathcal{F}_0 = \frac{1}{16\pi^2}\int_0^{\inf} d^2\r
\oint_\gamma\frac{dk}{2\pi i} \ln(\r^2+k^2)\frac{Q'(k)}{Q(k)}, 
\end{equation} i.e.\ the
equation (\ref{contour}) in section~\ref{analytics}.

For a uniform system in a periodic box of size $L$ the eigenvalues are $q_n =
\frac{2 n \pi}{L}$, so we can choose for example\footnote{Notice that this
  functions has zeros of order $p=2$. This introduces the same extra factor
  $1/2$ as in eq.~(\ref{Tr_ln_app}).}
\begin{equation}
Q_0(q)=\cos(q L)-1
\end{equation}
For the single-slab system considered in section~\ref{analytics}, the
eigenvalue equation (\ref{eigen_eq}) fixes $\tilde{Q}(q)$ up to an overall
constant, which can be in turn set by the requirement that $\tilde{Q}(q)$
reduces to $Q_0(q)$ when $\e_1 = \e_0$.
\begin{equation}
\tilde{Q}(q)=\left({\e_1-\e_0\over \e_1+\e_0}\right)^2 
\left(1-\cos[q(2a - L)]\right)- \left(1-\cos[q L]\right).
\end{equation}
The resulting spectral function, 
\begin{equation}
Q(q) = \tilde{Q}(q)/Q_0(q) =
 1- \left(\frac{\e_1-\e_0}{\e_1+\e_0}\right)^2\frac{1-\cos(q(L-2a))}{1-\cos(qL)},
\end{equation}
coincides with the one given in eq. (\ref{Q}), which as shown in
section~\ref{analytics}, when used in eq.~(\ref{contour_app}) gives a finite
result for the surface free energy, up to a (sub-leading) divergence in the
surface tension which will be discussed below.

%\subsection{Surface Self-Energy}
In section IV we found that we had to subtract an additional divergence that
can be traced to the integration constant introduced in eq. (\ref{fbar}). This
originates from eq. (\ref{contour_app}) being still ill-defined; only its
derivative with respect to $\rho$ is finite. In general however, there are
situations when one can evaluate eq. (\ref{contour_app}) directly and obtain a
finite result. This is the case when when $Q'(k)/Q(k)$ vanishes sufficiently
fast for large $|k|$. Under this assumption, let us go back to the contour
integration along the path shown in fig. 3 (a); we can neglect the integration
over the large circle already at the level of eq. (\ref{contour_app}), and
integrate by parts in the remaining contribution along the cuts
\begin{equation} \label{cont_int}
f(\r)=\oint_{cut}\frac{dk}{2\pi i}
 \ln[(\r^2+k^2)]\frac{Q'(k)}{Q(k)}=
 -\oint_{cut}\frac{dk}{2\pi i}\ln Q(k)\;\frac{2k}{\r^2+k^2},
\end{equation}
At this point, the branch
cuts $(-i\infty,-i\r)$ and $(i\r,-i\infty)$ have disappeared, and we can again
deform the contour to the two circles enclosing the two poles in $\pm i\r$, as
shown in figure~\ref{fig_2} (b). Then the residue theorem gives
\begin{equation}
f(\r) = 2\ln Q(i\r), 
\end{equation}
which leads to the finite result, eq. (\ref{F_general}). Notice that this
procedure fails in the  example of section~IV, since $Q(k)$ defined in
eq. (\ref{Q}) does not have the required property for large $|k|$.
  
The key observation is that, as we showed earlier in this appendix, 
$Q(k)$ must be defined in such a way as to subtract the
leading vacuum energy divergence, so that
\begin{equation}
Q(k) = {\tilde{Q}(k) \over Q_0(k)}, 
\end{equation}
where $Q_0(k)$ is a function giving the vacuum eigenvalue distribution, and
$\tilde{Q}(k)$ is such that firstly it gives the eigenvalue distribution of
the system under consideration and secondly it reduces to $Q_0(k)$ for a
uniform system.  In all physical situations, \emph{the large momentum modes
  should always behave like in the vacuum beyond a certain threshold},
determined by the properties of the material under investigation.  Therefore
it is natural to assume that in all physical systems, $Q(k) \to 1$ as $|k| \to
\infty$, which automatically guarantees that $Q'(k)/Q(k)\to 0$ for large
$|k|$.  In the case of a single plane slab, it is unphysical to assume that
the dielectric constant changes discontinuously, because that would imply that
\emph{all modes}, with arbitrary short wavelength, are affected by the
presence of the interface, as one can see from the matching conditions
(\ref{psi5}).  It is more reasonable to assume that modes with wavelengths
shorter than a certain cut-off $\delta$ (the molecular or atomic scale of the
dielectric) will not be sensitive to the difference between the dielectric and
the vacuum, therefore $Q(k)\sim 1$ for modes with $|k|> 1/\delta $.  Following
these considerations it is clear that the divergence in the surface
self-energy in section~IV is exclusively related to the singular nature of the
background under consideration, and that in a physical system with a smooth
dielectric function $\e({\bf r})$ the free energy in eq.  (\ref{free_surf}) is
finite.

\bibliography{thermalC_PRE}{}

\begin{thebibliography}{25}
\expandafter\ifx\csname natexlab\endcsname\relax\def\natexlab#1{#1}\fi
\expandafter\ifx\csname bibnamefont\endcsname\relax
  \def\bibnamefont#1{#1}\fi
\expandafter\ifx\csname bibfnamefont\endcsname\relax
  \def\bibfnamefont#1{#1}\fi
\expandafter\ifx\csname citenamefont\endcsname\relax
  \def\citenamefont#1{#1}\fi
\expandafter\ifx\csname url\endcsname\relax
  \def\url#1{\texttt{#1}}\fi
\expandafter\ifx\csname urlprefix\endcsname\relax\def\urlprefix{URL }\fi
\providecommand{\bibinfo}[2]{#2}
\providecommand{\eprint}[2][]{\url{#2}}

\bibitem[{\citenamefont{Casimir}(1948)}]{Casimir}
\bibinfo{author}{\bibfnamefont{H.~B.~K.} \bibnamefont{Casimir}},
  \bibinfo{journal}{Proc.Kon.Ned.Wet} \textbf{\bibinfo{volume}{51}},
  \bibinfo{pages}{793} (\bibinfo{year}{1948}).

\bibitem[{\citenamefont{Dzyaloshinskii
  et~al.}(1961)\citenamefont{Dzyaloshinskii, Lifshitz, and
  Pitaevskii}}]{Lifshitz}
\bibinfo{author}{\bibfnamefont{I.~E.} \bibnamefont{Dzyaloshinskii}},
  \bibinfo{author}{\bibfnamefont{E.~M.} \bibnamefont{Lifshitz}},
  \bibnamefont{and} \bibinfo{author}{\bibfnamefont{L.~P.}
  \bibnamefont{Pitaevskii}}, \bibinfo{journal}{Adv. Phys.}
  \textbf{\bibinfo{volume}{10}}, \bibinfo{pages}{165} (\bibinfo{year}{1961}).

\bibitem[{\citenamefont{Israelashvili and Tabor}(1973)}]{Tabor}
\bibinfo{author}{\bibfnamefont{J.~N.} \bibnamefont{Israelashvili}}
  \bibnamefont{and} \bibinfo{author}{\bibfnamefont{D.}~\bibnamefont{Tabor}},
  \bibinfo{journal}{Prog. Surface Membrane Sci.} \textbf{\bibinfo{volume}{7}},
  \bibinfo{pages}{1} (\bibinfo{year}{1973}).

\bibitem[{\citenamefont{Milton}(2004)}]{Milton}
\bibinfo{author}{\bibfnamefont{K.~A.} \bibnamefont{Milton}},
  \bibinfo{journal}{Journal of Physics A} \textbf{\bibinfo{volume}{37}},
  \bibinfo{pages}{R209} (\bibinfo{year}{2004}).

\bibitem[{\citenamefont{Ninham and Parsegian}(1970{\natexlab{a}})}]{Parsegian2}
\bibinfo{author}{\bibfnamefont{B.}~\bibnamefont{Ninham}} \bibnamefont{and}
  \bibinfo{author}{\bibfnamefont{V.}~\bibnamefont{Parsegian}},
  \bibinfo{journal}{Biophysical Journal} \textbf{\bibinfo{volume}{10}},
  \bibinfo{pages}{646} (\bibinfo{year}{1970}{\natexlab{a}}).

\bibitem[{\citenamefont{Parsegian and Ninham}(1970)}]{Parsegian}
\bibinfo{author}{\bibfnamefont{V.}~\bibnamefont{Parsegian}} \bibnamefont{and}
  \bibinfo{author}{\bibfnamefont{B.}~\bibnamefont{Ninham}},
  \bibinfo{journal}{Biophysical Journal} \textbf{\bibinfo{volume}{10}},
  \bibinfo{pages}{664} (\bibinfo{year}{1970}).

\bibitem[{\citenamefont{Netz}(2001)}]{Netz}
\bibinfo{author}{\bibfnamefont{R.}~\bibnamefont{Netz}}, \bibinfo{journal}{Eur.
  Phys. J. E.} \textbf{\bibinfo{volume}{5}}, \bibinfo{pages}{189}
  (\bibinfo{year}{2001}).

\bibitem[{\citenamefont{Dean and Horgan}(2005)}]{Dean1}
\bibinfo{author}{\bibfnamefont{D.~S.} \bibnamefont{Dean}} \bibnamefont{and}
  \bibinfo{author}{\bibfnamefont{R.~R.} \bibnamefont{Horgan}},
  \bibinfo{journal}{Physical Review E} \textbf{\bibinfo{volume}{71}},
  \bibinfo{pages}{041907} (\bibinfo{year}{2005}).

\bibitem[{\citenamefont{Kampen et~al.}(1968)\citenamefont{Kampen, Nijboer, and
  Schram}}]{Kampen}
\bibinfo{author}{\bibfnamefont{N.~G.~V.} \bibnamefont{Kampen}},
  \bibinfo{author}{\bibfnamefont{B.~R.} \bibnamefont{Nijboer}},
  \bibnamefont{and} \bibinfo{author}{\bibfnamefont{K.}~\bibnamefont{Schram}},
  \bibinfo{journal}{Physics Letters} \textbf{\bibinfo{volume}{26A}},
  \bibinfo{pages}{307} (\bibinfo{year}{1968}).

\bibitem[{\citenamefont{Ninham and Parsegian}(1970{\natexlab{b}})}]{Ninham}
\bibinfo{author}{\bibfnamefont{B.~W.} \bibnamefont{Ninham}} \bibnamefont{and}
  \bibinfo{author}{\bibfnamefont{V.~A.} \bibnamefont{Parsegian}},
  \bibinfo{journal}{Journal of Chemical physics} \textbf{\bibinfo{volume}{52}},
  \bibinfo{pages}{4578} (\bibinfo{year}{1970}{\natexlab{b}}).

\bibitem[{\citenamefont{Buescher and Emig}(2004)}]{Emig1}
\bibinfo{author}{\bibfnamefont{R.}~\bibnamefont{Buescher}} \bibnamefont{and}
  \bibinfo{author}{\bibfnamefont{T.}~\bibnamefont{Emig}},
  \bibinfo{journal}{Physical Review A} \textbf{\bibinfo{volume}{69}},
  \bibinfo{pages}{062101} (\bibinfo{year}{2004}).

\bibitem[{\citenamefont{Gies and Klingm\"uller}(2006)}]{Gies}
\bibinfo{author}{\bibfnamefont{H.}~\bibnamefont{Gies}} \bibnamefont{and}
  \bibinfo{author}{\bibfnamefont{K.}~\bibnamefont{Klingm\"uller}},
  \bibinfo{journal}{PRL} \textbf{\bibinfo{volume}{96}}, \bibinfo{pages}{220401}
  (\bibinfo{year}{2006}).

\bibitem[{\citenamefont{Feinberg et~al.}(2001)\citenamefont{Feinberg, Mann, and
  Revzen}}]{Feinberg}
\bibinfo{author}{\bibfnamefont{J.}~\bibnamefont{Feinberg}},
  \bibinfo{author}{\bibfnamefont{A.}~\bibnamefont{Mann}}, \bibnamefont{and}
  \bibinfo{author}{\bibfnamefont{M.}~\bibnamefont{Revzen}},
  \bibinfo{journal}{Annals of Physics} \textbf{\bibinfo{volume}{288}},
  \bibinfo{pages}{103} (\bibinfo{year}{2001}).

\bibitem[{\citenamefont{Pasquali and Maggs}()}]{PRL}
\bibinfo{author}{\bibfnamefont{S.}~\bibnamefont{Pasquali}} \bibnamefont{and}
  \bibinfo{author}{\bibfnamefont{A.~C.} \bibnamefont{Maggs}},
  \urlprefix\url{http://xxx.lanl.gov/pdf/0704.2171}.

\bibitem[{\citenamefont{Ninham and Daicic}(1998)}]{Daicic}
\bibinfo{author}{\bibfnamefont{B.~W.} \bibnamefont{Ninham}} \bibnamefont{and}
  \bibinfo{author}{\bibfnamefont{J.}~\bibnamefont{Daicic}},
  \bibinfo{journal}{Phys. Rev. A} \textbf{\bibinfo{volume}{57}},
  \bibinfo{pages}{1870} (\bibinfo{year}{1998}).

\bibitem[{\citenamefont{Maggs and Everaers}(2006)}]{Tony-Ralph}
\bibinfo{author}{\bibfnamefont{A.~C.} \bibnamefont{Maggs}} \bibnamefont{and}
  \bibinfo{author}{\bibfnamefont{R.}~\bibnamefont{Everaers}},
  \bibinfo{journal}{Physical Review Letters} \textbf{\bibinfo{volume}{96}},
  \bibinfo{pages}{230603} (\bibinfo{year}{2006}).

\bibitem[{\citenamefont{Maggs and Rossetto}(2002)}]{Tony}
\bibinfo{author}{\bibfnamefont{A.~C.} \bibnamefont{Maggs}} \bibnamefont{and}
  \bibinfo{author}{\bibfnamefont{V.}~\bibnamefont{Rossetto}},
  \bibinfo{journal}{Physical Review Letters} \textbf{\bibinfo{volume}{88}},
  \bibinfo{pages}{196402} (\bibinfo{year}{2002}).

\bibitem[{\citenamefont{Hertzberg et~al.}(2005)\citenamefont{Hertzberg, Jaffe,
  Kardar, and Scardicchio}}]{Jaffe1}
\bibinfo{author}{\bibfnamefont{M.~P.} \bibnamefont{Hertzberg}},
  \bibinfo{author}{\bibfnamefont{R.~L.} \bibnamefont{Jaffe}},
  \bibinfo{author}{\bibfnamefont{M.}~\bibnamefont{Kardar}}, \bibnamefont{and}
  \bibinfo{author}{\bibfnamefont{A.}~\bibnamefont{Scardicchio}},
  \bibinfo{journal}{Physical Review Letters} \textbf{\bibinfo{volume}{95}},
  \bibinfo{pages}{250402} (\bibinfo{year}{2005}).

\bibitem[{\citenamefont{Jaffe}(2003)}]{Jaffe2}
\bibinfo{author}{\bibfnamefont{R.~L.} \bibnamefont{Jaffe}},
  \emph{\bibinfo{title}{Unnatural acts: Unphysical consequences of imposing
  boundary conditions on quantum fields}} (\bibinfo{year}{2003}),
  \urlprefix\url{http://www.citebase.org/abstract?id=oai:arXiv.org:hep-th/0307%
014}.

\bibitem[{\citenamefont{Pasichnyk et~al.}(2007)\citenamefont{Pasichnyk,
  Everaers, and Maggs}}]{Igor}
\bibinfo{author}{\bibfnamefont{I.}~\bibnamefont{Pasichnyk}},
  \bibinfo{author}{\bibfnamefont{R.}~\bibnamefont{Everaers}}, \bibnamefont{and}
  \bibinfo{author}{\bibfnamefont{A.}~\bibnamefont{Maggs}},
  \bibinfo{journal}{submitted}  (\bibinfo{year}{2007}),
  \urlprefix\url{http://www.pct.espci.fr/~tony/lifshitz/igor.pdf}.

\bibitem[{\citenamefont{Levrel and Maggs}(2005)}]{MaggsW}
\bibinfo{author}{\bibfnamefont{L.}~\bibnamefont{Levrel}} \bibnamefont{and}
  \bibinfo{author}{\bibfnamefont{A.~C.} \bibnamefont{Maggs}},
  \bibinfo{journal}{Physical Review E} \textbf{\bibinfo{volume}{72}},
  \bibinfo{pages}{016715} (\bibinfo{year}{2005}).

\bibitem[{\citenamefont{Ninham and Parsegian}(1970{\natexlab{c}})}]{Ninham2}
\bibinfo{author}{\bibfnamefont{B.~W.} \bibnamefont{Ninham}} \bibnamefont{and}
  \bibinfo{author}{\bibfnamefont{V.~A.} \bibnamefont{Parsegian}},
  \bibinfo{journal}{Journal of Chemical Physics} \textbf{\bibinfo{volume}{53}},
  \bibinfo{pages}{3398} (\bibinfo{year}{1970}{\natexlab{c}}).

\bibitem[{\citenamefont{Emig et~al.}(2003)\citenamefont{Emig, Hanke,
  Golestanian, and Kardar}}]{Emig2}
\bibinfo{author}{\bibfnamefont{T.}~\bibnamefont{Emig}},
  \bibinfo{author}{\bibfnamefont{A.}~\bibnamefont{Hanke}},
  \bibinfo{author}{\bibfnamefont{R.}~\bibnamefont{Golestanian}},
  \bibnamefont{and} \bibinfo{author}{\bibfnamefont{M.}~\bibnamefont{Kardar}},
  \bibinfo{journal}{Physical Review A} \textbf{\bibinfo{volume}{67}},
  \bibinfo{pages}{022114} (\bibinfo{year}{2003}).

\bibitem[{\citenamefont{Chen et~al.}(2002)\citenamefont{Chen, Mohideen,
  Klimchitskaya, and Mostepanenko}}]{Chen}
\bibinfo{author}{\bibfnamefont{F.}~\bibnamefont{Chen}},
  \bibinfo{author}{\bibfnamefont{U.}~\bibnamefont{Mohideen}},
  \bibinfo{author}{\bibfnamefont{G.~L.} \bibnamefont{Klimchitskaya}},
  \bibnamefont{and} \bibinfo{author}{\bibfnamefont{V.~M.}
  \bibnamefont{Mostepanenko}}, \bibinfo{journal}{Physical Review Letters}
  \textbf{\bibinfo{volume}{88}}, \bibinfo{pages}{101801}
  (\bibinfo{year}{2002}).

\bibitem[{\citenamefont{Emig}(2003)}]{Emig3}
\bibinfo{author}{\bibfnamefont{T.}~\bibnamefont{Emig}},
  \bibinfo{journal}{Europhysics letters} \textbf{\bibinfo{volume}{62}},
  \bibinfo{pages}{466} (\bibinfo{year}{2003}).

\end{thebibliography}
\end{document}